\documentclass[11pt]{article}
\usepackage[utf8]{inputenc}
\usepackage{amsfonts}
\usepackage{amsmath}
\usepackage{amssymb}
\usepackage{indentfirst}
\usepackage{graphicx}
\usepackage[colorlinks]{hyperref}
\usepackage{cite}
\usepackage{array}
\usepackage{microtype}
\usepackage{soul}
\usepackage[normalem]{ulem}
\usepackage{cancel}
\usepackage{soul}
\usepackage{xspace}

\numberwithin{equation}{section}
\allowdisplaybreaks

\makeatletter

\begin{document}

\begin{titlepage}
\vspace{3cm}

\baselineskip=24pt

\begin{center}
\textbf{\LARGE Non-relativistic limit of the Mielke-Baekler gravity theory}
\par\end{center}{\LARGE \par}

\begin{center}
	\vspace{1cm}
	\textbf{Patrick Concha}$^{\ast,\star}$,
	\textbf{Nelson Merino}$^{\dag,\ddag}$,
	\textbf{Evelyn Rodríguez}$^{\ast,\star}$,
	\small
	\\[5mm]
    $^{\ast}$\textit{Departamento de Matemática y Física Aplicadas, }\\
	\textit{ Universidad Católica de la Santísima Concepción, }\\
\textit{ Alonso de Ribera 2850, Concepción, Chile.}
    \\[2mm]
    $^{\star}$\textit{Grupo de Investigación en Física Teórica, GIFT, }\\
	\textit{Concepción, Chile.}
\\[2mm]
    $^{\dag}$\textit{Instituto de Ciencias Exactas y Naturales,}\\\textit{Universidad Arturo Prat, Playa Brava 3265, 1111346, Iquique, Chile.}
    \\[2mm]
    $^{\ddag}$\textit{Facultad de Ciencias, Universidad Arturo Prat,}\\\textit{Avenida Arturo Prat Chacón 2120, 1110939, Iquique, Chile.}
	 \\[5mm]
	\footnotesize
	\texttt{patrick.concha@ucsc.cl},
	\texttt{nemerino@unap.cl},
	\texttt{erodriguez@ucsc.cl},
	\par\end{center}
\vskip 26pt
\begin{abstract}

In this paper, we present the most general non-relativistic Chern-Simons gravity model in three spacetime dimensions. We first study the non-relativistic limit of the Mielke-Baekler gravity through a contraction process. The resulting non-relativistic theory contains a source for the spatial component of the torsion and the curvature measured in terms of two parameters, denoted by $p$ and $q$.  We then extend our results by defining a Newtonian version of the Mielke-Baekler gravity theory, based on a Newtonian like algebra which is obtained from the non-relativistic limit of an enhanced and enlarged relativistic algebra. Remarkably, in both cases, different known non-relativistic and Newtonian gravity theories can be derived by fixing the $\left(p,q\right)$ parameters. In particular, torsionless models are recovered for $q=0$.

\end{abstract}
\end{titlepage}\newpage {} 

{\baselineskip=12pt \tableofcontents{}}

\section{Introduction}

Three-dimensional gravity theory has proven to be an interesting laboratory to study diverse aspects and features of the gravitational interaction and the underlying laws of quantum gravity. They share many properties with higher-dimensional gravity theories, as the existence of black hole solutions and their thermodynamical behavior \cite{Banados:1992wn,Banados:1998ta,Carlip:2005zn}. Moreover, three-dimensional gravity models admit non-perturbative quantizations \cite{Witten:1988hc}, possess rich and non-trivial boundary dynamics after considering suitable boundary conditions \cite{Brown:1986nw} and offer us a consistent way of coupling gravity with higher-spin gauge fields \cite{Aragone:1983sz,Campoleoni:2010zq,Fuentealba:2015jma}, among others.

The most general gravity Lagrangian being Lorentz-invariant with both curvature and torsion can be formulated through the Mielke-Baekler (MB) gravity Lagrangian \cite{Mielke:1991nn, Baekler:1992ab}. The MB gravity theory contains the usual Einstein-Hilbert gravity term, a cosmological constant term plus translational and rotational terms, each one with independent coupling constants. As a consequence, this theory is characterized by containing a source for both the constant Lorentz curvature and the constant torsion measured by parameters $p$ and $q$, respectively. Remarkably, particular choices of the parameters $\left(p,q\right)$ allow us to reproduce the usual Einstein-Hilbert gravity Lagrangian, the Teleparallel gravity and the "exotic" Witten Lagrangian \cite{Witten:1988hc}. Thus, the MB gravity is a useful toy model to study and analyze the role of the torsion and curvature in the AdS/CFT correspondence \cite{Maldacena:1997re}. Different issues about the MB gravity were explored in \cite{Blagojevic:2003uc,Blagojevic:2003vn,Blagojevic:2003wn,Blagojevic:2006hh,Blagojevic:2006jk,Blagojevic:2006nf,Cacciatori:2005wz,Giacomini:2006dr,Cvetkovic:2007sr,Klemm:2007yu,Santamaria:2011cz,Blagojevic:2013bu,Peleteiro:2020ubv,Geiller:2020edh,Caroca:2021njq}, such as its relation with the Chern-Simons (CS) action, black hole solutions, asymptotic symmetries, holography, its supersymmetric extension and its coupling to higher-spin gauge fields.  

On the other hand, non-relativistic (NR) symmetries have received a growing interest due to their applications in strongly coupled condensed matter systems and non-relativistic effective field theories \cite{Son:2008ye,Balasubramanian:2008dm,Kachru:2008yh,Taylor:2008tg,Duval:2008jg,Bagchi:2009my,Hartnoll:2009sz,Bagchi:2009pe,Hoyos:2011ez,Son:2013rqa,Christensen:2013lma,Christensen:2013rfa,Abanov:2014ula,Hartong:2014oma,Hartong:2014pma,Hartong:2015wxa,Geracie:2015dea,Gromov:2015fda,Hartong:2015zia,Taylor:2015glc,Zaanen:2015oix,Devecioglu:2018apj}. In three spacetime dimensions, NR gravity can be formulated through a CS action under the so-called extended Bargmann algebra \cite{Papageorgiou:2009zc,Bergshoeff:2016lwr}. Such NR symmetry differs from the Galilei algebra by two central charges which are required to avoid degeneracy of the invariant bilinear trace. It is also worth mentioning that the extended Bargmann gravity can be obtained as an NR limit of a U(1)-enlargement of the Poincaré CS gravity theory. In like manner, the inclusion of a cosmological constant requires considering the extended Newton-Hooke symmetry \cite{Aldrovandi:1998im,Gibbons:2003rv,Brugues:2006yd,Alvarez:2007fw,Papageorgiou:2010ud,Duval:2011mi,Hartong:2016yrf,Duval:2016tzi} which turns out to be the NR limit of the $AdS\oplus \mathfrak{u}(1)^2$ algebra. 

The accommodation of a non-vanishing torsion in an NR environment requires a more subtle treatment \cite{Bergshoeff:2015ija,Bergshoeff:2017dqq,VandenBleeken:2017rij}. Indeed Torsional Newton-Cartan gravity appears by gauging the Schrödinger algebra being the conformal extension of the Bargmann algebra \cite{Bergshoeff:2014uea}. In such formalism, the time-component of the torsion is non zero. Non-vanishing torsion condition has first been encountered in the context of Lifshitz holography \cite{Christensen:2013lma} and Quantum Hall Effect \cite{Geracie:2015dea}. An alternative approach has then been presented in \cite{Concha:2021llq,Concha:2022you}, considering the CS action for an NR teleparallel symmetry. In such a case, the cosmological constant appears as a source of the space-component of the torsion, while the time-component is zero. Although both formalisms are quite different, there are preliminary results in \cite{Concha:2021llq} that could reveal some relations between them.

To our knowledge, although the presence of torsion and curvature in NR gravity have been approached separately, there is no an MB analog in the NR regime. In this work, we present the NR counterpart of the MB gravity Lagrangian by applying a proper NR limit to the algebra introduced in \cite{Geiller:2020edh} which reproduces the MB Lagrangian using the CS form. To this end, we add two $\mathfrak{u}\left(1\right)$ generators which will provide us the desired quantities of central charges in the NR regime allowing us to avoid degeneracy. We show that, as in the relativistic MB model, the NR MB theory contains a source for the spatial component of both torsion and curvature depending on the parameters $q$ and $p$, respectively. Interestingly,
we recover different known NR gravity theories for particular values of the $\left(p,q\right)$ parameters. We then extend our results to a Newtonian generalization of the MB gravity considering the NR limit of an enhanced algebra. Diverse extended Newtonian gravity theories can also be recovered for particular choices of the $\left(p,q\right)$ parameters. In particular, the Newtonian MB algebra obtained here can be seen as a central extension of the Newtonian symmetry introduced in \cite{Hansen:2019pkl} in which an action principle for Newtonian gravity has been derived in four spacetime dimensions.

The paper is organized as follows: In Section \ref{sec2} we briefly review the MB gravity theory and its CS formulation. Section \ref{sec3} and \ref{sec4} contain our main results. In Section \ref{sec3} we first present the NR version of the MB gravity theory by applying an NR limit to a $U\left(1\right)$-enlargement of the so-called MB algebra. Section \ref{sec4} is devoted to the construction of a Newtonian version of the MB gravity theory. Section \ref{sec5} concludes our work with some discussions about future developments.

\section{Mielke-Baekler Lagrangian and Chern-Simons formulation}\label{sec2}
A three-dimensional gravity model that is characterized by a non-vanishing torsion was proposed by Mielke and Baekler in \cite{Mielke:1991nn}.  The MB Lagrangian is the most general three-form constructed with the dreibein one-form $E^{A}$ and the dual spin connection one-form $W^{A}$, and reads as follows
\begin{equation}\label{MB}
    L_{\text{MB}}[E^{A},W^{A}]=\sigma_{0}L_{0}[E^{A}]+\sigma_{1} L_{1}[E^{A},W^{A}]+\sigma_{2} L_{2}[W^{A}]+\sigma_{3} L_{3}[E^{A},W^{A}]\,,
\end{equation}
where $\sigma_{i}, i=0,\dots,3$ are independent constants and
\begin{eqnarray}
L_{0}[E^{A}] & = & \frac{1}{3} \epsilon_{ABC}E^{A}E^{B}E^{C}\,,\nonumber\\
L_{1} [E^{A},W^{A}]& = &2 E_{A}R^{A}\,,\nonumber\\
L_{2} [W^{A}]& = &  W^A d W_A + \frac{1}{3}\epsilon^{ABC}W_A W_B W_C\,,\label{MBterms}\\
L_{3} [E^{A},W^{A}]& = & E_{A}T^{A}\,.\nonumber
\end{eqnarray}
Here
\begin{align*}
  R^{A}&=d W^{A}+\frac{1}{2}\epsilon^{ABC}W_{B}W_{C}\,,\\
  T^{A}&=dE^{A}+\epsilon^{ABC}W_{B}E_{C}\,,
\end{align*}
are the corresponding Lorentz curvature and torsion two-forms, $A=0,1,2$ are Lorentz indices which are lowered and raised with the Minkowski metric $\eta_{AB}=\left(-1,1,1\right)$ and $\epsilon_{ABC}$ is the three-dimensional Levi Civita tensor which satisfies $\epsilon_{012}=-\epsilon^{012}=1$. $L_0$ yields a cosmological constant term, $L_1$ corresponds to the Einstein-Hilbert Lagrangian, $L_2$ is the Chern-Simons gravitational term and $L_4$ represents a translational Chern-Simons term. For a particular choice of the $\sigma_{i}$ parameters, we can recover the Einstein-Hilbert gravity, teleparallel gravity and the "exotic" Witten's gravity \cite{Witten:1988hc}.

The equations of motion obtained from the MB Lagrangian are given by
\begin{align*}
   2\sigma_{1}R^{A}+\sigma_{0}\epsilon^{ABC}E_{B}E_{C}+2\sigma_{3}T^{A}&=0\,,\\
   2\sigma_{1}T^{A}+2\sigma_{2}R^{A}+\sigma_{3}\epsilon^{ABC}E_{B}E_{C}&=0\,.
\end{align*}
Then, assuming $ \sigma_{1}^2-\sigma_{2}\sigma_{3}\neq0$, the field equations can be rewritten as
\begin{equation}
    2T^A+q \epsilon^{ABC}E_{B}E_{C}=0\,, \qquad 2R^{A}+p\epsilon^{ABC}E_{B}E_{C}=0\,,\label{MBcurv1}
\end{equation}
where
\begin{equation*}
q:=\frac{\sigma_{1}\sigma_{3}-\sigma_{0}\sigma_{2}}{\sigma_{1}^{2}-\sigma_{2}\sigma_{3}}\,, \qquad
p:=\frac{\sigma_{0}\sigma_{1}-\sigma_{3}^2}{\sigma_{1}^{2}-\sigma_{2}\sigma_{3}}\,.
\end{equation*}
In this way, the field configurations are characterized by constant curvature and constant torsion. As we have mentioned before, we can identify three particular cases. First, the EH gravity with cosmological constant is recovered for $\sigma_{2}=\sigma_{3}=0$, such that the torsion vanishes ($q=0$). Second, the teleparallel gravity theory in three-dimensions, having a non-zero torsion and a vanishing curvature ($p=0$) can be obtained by fixing $\sigma_{0}\sigma_{1}-\sigma_{3}^2=0$. In such a case, the cosmological constant can be seen as a source for the torsion. Finally, considering $\sigma_{0}=\sigma_{1}=0$ we recover the Witten's exotic gravity \cite{Witten:1988hc}.

Let us note that the Riemann-Cartan curvature $R^{A}$ can be expressed in terms of its Riemannian part $\tilde R^{A}$ and the contorsion one-form $K^{A}$. Indeed, decomposing the dual spin-connection as $W^{A}=\tilde W^{A}+K^{A}$, where $\tilde W^{A}$ is the (torsionless) dual Levi-Civita connection, the curvature and torsion two-forms corresponding to the dual Levi-Civita connection are
\begin{equation}\label{Riem}
2\tilde R^A=\Lambda \epsilon^{ABC}E_B E_C\,,\qquad 2\tilde T^A=0
\end{equation}
where 
\begin{equation}
    \Lambda:=-\left(p+\frac{q^{2}}{4}\right)\,.\label{Lambda}
\end{equation}
Therefore, the MB gravity model describes constant curvature spacetimes with cosmological constant $\Lambda$.

\subsection{Chern-Simons formulation}
As it was shown in \cite{Blagojevic:2003vn,Cacciatori:2005wz, Giacomini:2006dr,Geiller:2020edh} the MB model can be written as a CS theory. Here, following \cite{Geiller:2020edh}, we will consider a CS formulation of the MB model in a particular basis which puts forward more clearly the role of the constants $\sigma_i$ and of the curvature and torsion parameters $(p,q)$. Let us consider the algebra spanned by generators $(J_{A},P_{A})$ which satisfy the following commutation relations
\begin{equation}\label{algebra01} 
\begin{split}
 &[J_A,J_B]=\epsilon_{ABC} J^{C} \,, \\
 &[J_A,P_B]=\epsilon_{ABC}P^{C} \,,\\
 &[P_A,P_B]=\epsilon_{ABC} \left(p J^{C}+q P^{C}\right) \,.
\end{split}
\end{equation}
Here $(p,q)$ can be arbitrary for the moment. As it was noticed in \cite{Geiller:2020edh}, the above algebra is actually isomorphic to the AdS algebra. Indeed, defining the new generators
\begin{equation}
    \hat{P}_{A}=P_{A}-\frac{q}{2}J_{A}\,,
\end{equation}
the algebra \eqref{algebra01} maps to
\begin{equation}\label{algebra02} 
\begin{split}
 &[J_A,J_B]=\epsilon_{ABC} J^{C} \,, \\
 &[J_A,\hat{P}_B]=\epsilon_{ABC}\hat{P}^{C} \,,\\
 &[\hat{P}_A,\hat{P}_B]=-\Lambda\epsilon_{ABC} J^{C}  \,,
\end{split}
\end{equation}
where $\Lambda$ is defined in \eqref{Lambda}. Just for simplicity, along this work we will refer to the AdS algebra in the specific basis \eqref{algebra01} as the \textit{MB algebra}. 

The  gauge connection one-form $A$ for the MB algebra can be defined as follows,
\begin{equation}\label{oneform1}
    A=W^{A}J_{A}+E^{A}P_{A}\,.
\end{equation}
where $W^{A}$ is the one-form spin connection and $E^{A}$ is the dreibein. Thus the curvature two-form reads
\begin{equation}
F=\mathcal{R}^{A}\left(W\right)J_{A}+\mathcal{R}^{A}\left(E\right)P_{A}\,,\label{curvMB}
\end{equation}
where
\begin{align*}
 \mathcal{R}^{A}\left(W\right)&:=d W^{A}+\frac{1}{2}\epsilon^{ABC}W_{B}W_{C}+\frac{p}{2}\epsilon^{ABC}E_{B}E_{C}\,,\\
 \mathcal{R}^{A}\left(E\right)&:=dE^{A}+\epsilon^{ABC}W_{B}E_{C}+\frac{q}{2}\epsilon^{ABC}E_{B}E_{C}\,.
\end{align*}
Additionally, the MB algebra \eqref{algebra01} 
admits an invariant bilinear form with the following non-vanishing components
\begin{equation}\label{invtensor}
    \langle J_{A} J_{B} \rangle = \sigma_{2} \,\eta_{A B}\,,\qquad \langle J_{A} P_{B} \rangle = \sigma_{1} \,\eta_{A B}\,,\qquad \langle P_{A} P_{B} \rangle = \left(p \sigma_{2}+q\sigma_{1}\right)\eta_{A B}\,.
\end{equation}

Considering the gauge connection one-form \eqref{oneform1} and the non-vanishing
components of the invariant tensor \eqref{invtensor} in the three-dimensional CS Lagrangian,
\begin{equation}
L_{\text{CS}}[A]= \left\langle AdA+\frac{2}{3}A^{3}\right\rangle \,,
\label{CS}
\end{equation}
we find, modulo boundary terms, that 
\begin{equation}
    L_{\text{CS}}[A]=\left(p\sigma_{1}+q\sigma_{3}\right)L_{0}[E^{A}]+\sigma_{1} L_{1}[E^{A},W^{A}]+\sigma_{2} L_{2}[W^{A}]+\sigma_{3} L_{3}[E^{A},W^{A}]\,,
\end{equation}
where $L_{0},...,L_{3}$ are given in \eqref{MBterms} and where we have imposed
\begin{equation}
      \sigma_{3}=p \sigma_{2}+q \sigma_{1}\,.
\end{equation}
Then, by further imposing
\begin{equation}
    \sigma_{0}=p \sigma_{1}+q \sigma_{3}\,,
\end{equation}
we find that, up to boundary terms \cite{Geiller:2020edh}
\begin{equation}
 L_{\text{CS}}[A]=  L_{\text{MB}}[E^{A},W^{A}]\,.
\end{equation}
Thus, the MB Lagrangian is a CS theory for the connection \eqref{oneform1} and the algebra spanned by $(J_{A},P_{A})$ \eqref{algebra01}. Requiring the condition for having a non-degenerate invariant tensor, the field equations coming from the CS Lagrangian correspond to the vanishing of the components of the curvature two-form \eqref{curvMB}, which can be expressed as in \eqref{MBcurv1}.

In the following sections, we will analyze non-relativistic versions of the MB CS gravity theory previously introduced. We will study the NR limits through a contraction process, in which the speed of light is taken to infinity ($c\rightarrow\infty$). As it is well-known, taking this limit in the relativistic Lagrangian might lead to divergences and degeneracy. One way to avoid such difficulties is to add extra fields to the relativistic theory. Then, in our case, we will include two new extra fields in order to obtain finite CS Lagrangians, constructed from NR algebras with a non-degenerate bilinear form.  First, we will consider the NR limit to an enlargement of the MB gravity. In the second part of the work, we will show that a Newtonian version of the MB CS gravity theory can be constructed from the contraction of an enhancement and enlargement of the MB algebra. In both cases, we will decompose the $A$-index as follows:
\begin{equation}
    A\rightarrow\left(0,a\right), \qquad  a=1,2\,. \label{decom}
\end{equation}
Then, we will apply particular redefinitions to the corresponding relativistic algebras and we will take the NR limits in order to get its NR and Newtonian versions.  We will also consider the contraction at the level of the invariant tensors in order to construct the corresponding NR CS gravity actions.

\subsection{U(1) Enlargements}
As it was previously discussed, we will add two extra fields to the MB CS gravity theory to obtain a finite and non-degenerate NR Lagrangian after the contraction process. We include two $U(1)$ one-form gauge fields $y_{1}$ and $y_{2}$ in the one-form gauge connection \eqref{oneform1}:
\begin{equation}\label{oneform2}
    A=W^{A}J_{A}+E^{A}P_{A}+y_{1}Y_{1}+y_{2}Y_{2}\,,
\end{equation}
where the new Abelian generators satisfy the following non-vanishing invariant tensors
\begin{equation}
    \langle Y_{1}Y_{1}\rangle=\sigma_{3}=p\sigma_{2}+q\sigma_{1}\,, \qquad \langle Y_{1}Y_{2}\rangle=\sigma_{1}\,,\qquad \langle Y_{2}Y_{2}\rangle=\sigma_{2}\,.\label{intenu}
\end{equation}
Then, the non-zero components of the invariant tensor for the algebra  $(J_{A},P_{A})$ enlarged with two $U(1)$ generators are given by \eqref{invtensor} along with \eqref{intenu}.  The relativistic enlarged CS Lagrangian is written as
 \begin{align}
 L_{\text{CS}}^{U(1)}&=\sigma_{0}L_{0}[E^{A}]+\sigma_{1}\left( L_{1}[E^{A},W^{A}]+2y_{1}dy_{2}\right)+\sigma_{2}\left( L_{2}[W^{A}]+y_{2}dy_{2}\right) \notag \\
   & +\sigma_{3} \left(L_{3}[E^{A},W^{A}]+y_{1}dy_{1}\right)\,.\label{MBenl}
 \end{align}
In the next sections, we will show that the inclusion of these extra gauge fields in the MB CS theory is essential as they allow to cancel the divergences appearing in the limiting process. Let us note that the motivation to consider the contraction of the enlarged algebra $(J_{A},P_{A},Y_{1},Y_{2})$ is twofold. First, as we will see, its NR version admits a non-degenerate invariant tensor. Second, the NR MB CS Lagrangian leads to different known NR CS gravity theories when the $(p,q)$ parameters are set to particular values. It is important to mention that, as we shall see, the U$\left(1\right)$-enlargement is also required to approach the Newtonian regime.
\section{Non-relativistic MB gravity}\label{sec3}
In this section, we approach the construction of the NR version of the previously introduced MB CS gravity. To this purpose, we will first consider the NR limit to the algebra \eqref{algebra01} enlarged with two $U(1)$ generators. It is obtained by performing the indices decomposition \eqref{decom}, and subsequently performing an Inönü-Wigner contraction, for which we introduce the dimensionless parameter $\xi$. We define the contraction process through the identification of the relativistic generators with the NR generators as
\begin{eqnarray}
    J_{0}&=&  \frac{J}{2}+\xi^2 S \,,\qquad  J_{a}=\xi G_{a} \,,  \qquad Y_{2}= \frac{J}{2}-\xi^2 S\,, \notag\\
   {P}_{0}&=& \frac{H}{2\xi}+\xi M\,, \qquad P_{a}= P_{a}\,, \qquad Y_{1}=\frac{H}{2\xi}-\xi M\,,
\end{eqnarray}
along with the following scaling
\begin{equation}
    p\rightarrow \frac{p}{\xi^{2}}\,, \qquad q\rightarrow\frac{q}{\xi}\,.
\end{equation}
which is required to have a well-defined limit $\xi\rightarrow\infty$. Then, after applying the previous steps to \eqref{algebra01}, we get a NR version of the MB algebra:
\vspace{5mm}
\begin{eqnarray}
\left[ J,G_{a}\right] &=&\epsilon _{ab}{G}_{b}\,, \qquad \qquad \quad \, \,
\left[ J,P_{a}\right] =\epsilon _{ab}{P}_{b}\,, \qquad \qquad \qquad \qquad \quad \,
\left[G_{a},{P}_{b}\right] =-\epsilon _{ab}M\,,  \notag
\\
\left[ G_{a},G_{a}\right] &=&-\epsilon _{ab}S\,,\qquad \qquad \ \
\left[ P_{a},P_{b}\right] =-\epsilon _{ab}\left(p S+q M\right)\,,\qquad \qquad\ \,
\left[ H,G_{a}\right] =\epsilon _{ab}P_{b}\,,  \notag \\
\left[ H,P_{a}\right]& =&\epsilon _{ab}\left(pG_{b}+qP_{b}\right)\,,
  \label{MBNR}
\end{eqnarray}
where $a=1,2$, $\epsilon_{ab}\equiv\epsilon_{0ab}$,  $\epsilon^{ab}\equiv\epsilon^{0ab}$. This NR algebra consists of spatial
translations $P_a$, spatial rotations $J$, Galilean boosts $G_a$, time translations $H$ and two central charges $S$ and $M$. Let us note that different known NR algebras can be derived from \eqref{MBNR} when the $(p,q)$ parameters are fixed to certain values. Indeed, the extended Bargmann algebra \cite{Papageorgiou:2009zc,Bergshoeff:2016lwr}, the torsional NR algebra presented in \cite{Concha:2021llq} and the extended Newton-Hooke algebra \cite{Papageorgiou:2010ud,Duval:2011mi,Hartong:2016yrf,Duval:2016tzi} are obtained when the parameters are set as shown in Table \ref{Table1}.
\begin{table}[!h]
\centering
    \begin{tabular}{|c|c|c|}
\hline
 NR algebra  & $p $ & $q$ \\ \hline\hline
NR torsional algebra & 
   $0$ & $-2/\ell$ \\
   Extended Newton-Hooke algebra &
   $1/\ell^{2}$ & $0$ \\ 
   Extended Bargmann algebra & $0$ & $0$\\  \hline
\end{tabular}\notag
\caption{Non-relativistic symmetries for different values of $p$ and $q$ in the NR MB algebra.}
\label{Table1}
\end{table}
Let us note that the presence of the two central charges $S$ and $M$ is essential to have a non-degenerate invariant tensor.  Indeed, when we set $M=S=0$, the resulting algebra corresponds to the torsional galilean-AdS algebra introduced in \cite{Matulich:2019cdo}, which can not be equipped with a non-degenerate invariant bilinear form. In this way, the NR MB algebra \eqref{MBNR} can be seen as a central extension of the torsional galilean-AdS algebra.

\subsection{Non-relativistic Chern-Simons Lagrangian}
Now, we extend our study to the explicit construction of a CS action for the NR algebra \eqref{MBNR}.
To this end, let us consider the corresponding gauge connection one-form $A$, 
\begin{equation}
    A=\tau H+e^{a}P_{a}+\omega J+\omega^{a}G_{a}+mM+sS\,. \label{oneformMBNR}
\end{equation}
The curvature two-form $F=dA+\frac{1}{2}\left[ A,A\right] $ is given by%
\begin{eqnarray}
F =R\left(\tau\right) H+R^{a}\left(e^{b}\right)P_{a}+R\left(\omega\right) J+R^{a}\left(\omega^{b}\right)G_{a}+R\left(m\right)M+R\left(s\right)S\,, \label{curvMBnon}
\end{eqnarray}
where the components are explicitly given by:
\begin{align}
R\left( \tau \right) &=d\tau \,, &  R^{a}\left( e ^{b}\right)  &= de ^{a}+\epsilon^{ac}\omega e_{c}+\epsilon^{ac}\tau \omega_{c}+q\epsilon^{ac}\tau e_{c}\,,  \notag \\
R\left( \omega \right) &= d\omega \,, & R^{a}\left( \omega^{b}\right) &= d \omega^a + \epsilon ^{ac}\omega \omega_{c} +p \epsilon^{ac}\tau e_{c}\,,  \notag \\
R\left( m\right) &= dm+\epsilon^{ac}e_{a}\omega_{c}+\frac{q}{2}\epsilon^{ac}e_{a}e_{c}  \,, & R\left( s\right) &= ds+ \frac{1}{2}\epsilon ^{ac}\omega_{a} \omega_{c} + \frac{p}{2}\epsilon^{ac}e_{a}e_{c} \,.  \label{curvMCA}
\end{align}
Naturally, when we fix the $(p,q)$ parameters to the values given in the table, the NR two-form curvatures constructed from those algebras are recovered.

The non-vanishing components of a non-degenerate invariant tensor are obtained by applying the contraction process to the relativistic invariant tensors \eqref{invtensor} and \eqref{intenu}. These are given by
\begin{align}
\langle J S\rangle &=-\tilde{\sigma}_{2}\,,\notag\\
\langle G_{a} G_{b} \rangle &=\tilde{\sigma}_{2}\delta_{ab}\,, \notag\\
\langle G_{a} P_{b} \rangle &=\tilde{\sigma}_{1}\delta_{ab}\,, \notag\\
\langle H S \rangle &=\langle M J \rangle=-\tilde{\sigma}_{1}\,, \notag\\
\langle P_{a} P_{b} \rangle &=\left(p\tilde{\sigma}_{2}+q\tilde{\sigma}_{1}\right)\delta_{ab}\,, \notag\\
\langle H M \rangle &=-\left(p\tilde{\sigma}_{2}+q\tilde{\sigma}_{1}\right)\,,\label{ITNRMB}
\end{align}
where we have considered the following rescaling for the $\sigma_{1}$ and $\sigma_{2}$ parameters
\begin{equation}\label{re}
    \sigma_{1}=\tilde{\sigma}_{1}\xi\,, \qquad\sigma_{2}=\tilde{\sigma}_{2}\xi^{2}\,
\end{equation}
which is required to end with a finite NR CS Lagrangian. Then, the NR CS Lagrangian gauge-invariant under the NR MB algebra \eqref{MBNR} is
\begin{eqnarray}
L_{\text{NRMB}} &=& -\tilde{\sigma}_{0}\epsilon^{ac}\tau e_{a}e_{c}+\tilde{\sigma}_{1}\left[
e_{a}\hat{R}^{a}\left(\omega ^{b}\right)+\omega_{a}\hat{R}^{a}\left(e ^{b}\right)-2mR(\omega )-2s R(\tau)\right] \notag \\  
&& +\tilde{\sigma}_{2}\left[ \omega _{a}\hat{R}^{a}\left(\omega
^{b}\right)-2sR\left( \omega \right) \right]+\tilde{\sigma}_{3}\left[
e_{a}\hat{R}^{a}\left(e ^{b}\right)-mR(\tau )-\tau \hat{R}(m)\right] \,.  \label{NRCS}
\end{eqnarray}%
where we have defined 
\begin{align}
    \hat{R}^{a}\left( e ^{b}\right)  &= de ^{a}+\epsilon^{ac}\omega e_{c}+\epsilon^{ac}\tau \omega_{c}\,, \notag \\
     \hat{R}^{a}\left( \omega^{b}\right) &= d \omega^a + \epsilon ^{ac}\omega \omega_{c}\,,\\
    \hat{R}\left( m\right) &= dm+ \epsilon ^{ac}e_{a} \omega_{c}\,, 
\end{align}
and
\begin{align}
\tilde{\sigma}_{3}&=p \tilde{\sigma}_{2}+q \tilde{\sigma}_{1}\,, \notag \\
    \tilde{\sigma}_{0}&=p\tilde{\sigma}_{1}+q \tilde{\sigma}_{3} \label{s30}
\end{align}
The NR Lagrangian \eqref{NRCS} corresponds to the NR counterpart of the MB gravity Lagrangian \eqref{MB} and can be seen as the most general NR gravity Lagrangian in three-dimensions invariant under the NR version of the MB algebra \eqref{MBNR}. As Table \ref{Table1} indicates, depending on the values of $p$ and $q$, the previous Lagrangian leads to different NR gravity theories. When $p=q=0$, and consequently when $\tilde{\sigma}_{3}=\tilde{\sigma}_{0}=0$, the NR Lagrangian corresponds to the Extended Bargmann gravity \cite{Bergshoeff:2016lwr}. On the other hand, when $q=0$ and $p=1/\ell^{2}$, the Lagrangian reduces to the Extended Newton-Hooke gravity theory.  Setting $p=0$ and $q=-2/\ell$ the Lagrangian reproduces the NR torsional gravity introduced in \cite{Concha:2021llq}. For arbitrary values of $p$ and $q$, the equations of motion are given by the vanishing of the curvatures \eqref{curvMCA}. Fixing $p$ and $q$ as it was discussed above, reproduces diverse NR dynamics with and without non-zero spatial torsion $\hat{R}^{a}\left(e^{a}\right)$.

The non-degeneracy of the invariant bilinear form \eqref{ITNRMB} ensures that the action \eqref{NRCS} involves a kinetic term for each gauge field and the field equations of the theory
are given by the vanishing of the curvature two-form \eqref{curvMCA}. Indeed, the equations of motion derived from \eqref{NRCS} are given by
\begin{eqnarray}
    \delta\omega^{a}&:& \qquad \tilde{\sigma}_{1}R^{a}\left(e^{b}\right)+\tilde{\sigma}_{2}R^{a}\left(\omega^{b}\right)=0\,, \notag \\
    \delta\omega&:& \qquad \tilde{\sigma}_{1}R\left(m\right)+\tilde{\sigma}_{2}R\left(s\right)=0\,, \notag \\
    \delta e^{a}&:& \qquad \tilde{\sigma}_{1}\left[R^{a}\left(\omega^{b}\right)+qR^{a}\left(e^{b}\right)\right]+\tilde{\sigma}_{2}\,pR^{a}\left(e^{b}\right)=0\,, \notag \\
    \delta\tau&:& \qquad \tilde{\sigma}_{1}\left[R\left(s\right)+qR\left(m\right)\right]+\tilde{\sigma}_{2} \,
    pR\left(m\right)=0\,, \notag \\
    \delta s&:& \qquad \tilde{\sigma}_{1}R\left(\tau\right)+\tilde{\sigma}_{2}R\left(\omega\right)=0\,, \notag \\
    \delta m&:& \qquad \tilde{\sigma}_{1}\left[R\left(\omega\right)+qR\left(\tau\right)\right]+\tilde{\sigma}_{2}\,pR\left(\tau\right)=0\,, \label{eom1}
\end{eqnarray}
where we have used \eqref{s30}. In particular, the non-degeneracy of the invariant tensor \eqref{ITNRMB} is satisfied for $\tilde{\sigma}_1^{2}-\tilde{\sigma}_{2}\tilde{\sigma}_{3}\neq 0$ which implies the vanishing of the curvatures \eqref{curvMCA}. Of particular interest is the vanishing of $R^{a}\left(e^{b}\right)=0$ and $R^{a}\left(\omega^{b}\right)=0$ which implies the non-vanishing of the usual spatial torsion $\hat{R}^{a}\left(e^{b}\right)\neq0$ and the spatial curvature $\hat{R}^{a}\left(\omega^{b}\right)\neq0$. This feature is inherited from the relativistic MB theory which contains a source for both torsion and Lorentz curvature measured by a parameter $q$ and $p$, respectively. At the NR regime, the same behavior appears along the spatial component of the torsion $\hat{R}^{a}\left(e^{b}\right)=-q\epsilon^{ac}\tau e_c $ and the curvature $\hat{R}^{a}\left(\omega^{b}\right)=-p\epsilon^{ac}\tau e_c $. Naturally, for $q=0$, the theory is torsionless and the geometry remains Riemannian. It is also important to mention that there are no values for $\left(p,q\right)$ allowing to turn on the time component of the torsion $R\left(\tau\right)$ as in the torsional Newton-Cartan gravity \cite{Bergshoeff:2015ija,Bergshoeff:2017dqq,VandenBleeken:2017rij}. As it was discussed in \cite{Concha:2021llq}, although our NR model contains a zero time-like torsion $R\left(\tau\right)$, the presence of a non-vanishing spatial torsion in a CS formalism could be useful for introducing non-zero time-like torsion. Unlike our approach, the torsional Newton-Cartan model appears by gauging the conformal extension of the Bargmann algebra \cite{Bergshoeff:2014uea}.

Let us note that the NR gravity action \eqref{NRCS} can be alternatively recovered from the relativistic $U\left(1\right)$-enlarged MB CS action \eqref{MBenl}. Indeed, one can express the relativistic gauge fields in terms of the NR ones as follows:
\begin{eqnarray}
W^{0} &=&\omega +\frac{s}{2\xi ^{2}}\,,\text{\ \ \ \ \ }W^{a}=\frac{\omega
^{a}}{\xi }\,,\text{ \ \ \ \ \ \ \thinspace \thinspace \thinspace }y_{2}=\omega -%
\frac{s}{2\xi ^{2}}\,,  \notag \\
E^{0} &=&\xi \tau +\frac{m}{2\xi }\,,\text{ \ \ \ \ \thinspace \thinspace }%
E^{a}=e^{a}\,,\text{ \ \ \ \ \ \ \ \ }y_{1}=\xi \tau -\frac{m}{2\xi }\,, \label{rescgf}
\end{eqnarray}
The NR CS action \eqref{NRCS} is obtained considering these last expressions along with the rescaling of the relativistic parameters \eqref{re} on the relativistic CS action \eqref{MBenl} and then applying the limit $\xi\rightarrow\infty$.
\section{Newtonian MB gravity}\label{sec4}

In this section, we present a Newtonian version of the MB gravity theory in three dimensions, which is based on a novel non-relativistic algebra obtained as a contraction of an enhancement and enlargement of the MB algebra \eqref{algebra01}. The obtained Newtonian MB symmetry results to be a central extension of the Newtonian algebra appearing as the underlying symmetry of an action principle for Newtonian gravity \cite{Hansen:2019pkl}.

\subsection{Enhanced MB algebra and U(1)-enlargement}
An enhancement of the MB algebra with two additional generators $\lbrace S_A, L_A \rbrace$ satisfies the commutators \eqref{algebra01} along with the following ones:
\begin{eqnarray}
\left[J_A,S_B\right]&=&\epsilon_{ABC} S^{C} \,, \qquad \qquad
\left[J_A,L_B\right]=\epsilon_{ABC} L^{C} \,, \notag \\
\left[S_A,P_B\right]&=&\epsilon_{ABC} L^{C} \,, \qquad \qquad
\left[L_A,P_B\right]=\epsilon_{ABC}  (p S^{C}+q L^{C}) \,. \label{EnhMB} 
\end{eqnarray}
One can notice that such enhancement is isomorphic to the coadjoint AdS algebra defined in \cite{Bergshoeff:2020fiz}. In fact, by considering the redefinition
\begin{eqnarray}
    \hat{P}_{A}&=&P_{A}-\frac{q}{2}J_{A}\,,\notag\\
    \hat{L}_{A}&=&L_{A}-\frac{q}{2}S_{A}\,,\notag
\end{eqnarray}
the algebra satisfies the coadjoint AdS commutation relations:
\begin{align}
    \left[J_A,J_B\right]&=\epsilon_{ABC}J^{C}\,,&
    \left[J_A,\hat{P}_B\right]&=\epsilon_{ABC}\hat{P}^{C}\,,\notag\\
    \left[\hat{P}_{A},\hat{P}_{B}\right]&=-\Lambda\epsilon_{ABC}J^{C}\,,&
    \left[J_A,S_B\right]&=\epsilon_{ABC}S^{C}\,,\notag\\
    \left[J_A,\hat{L}_B\right]&=\epsilon_{ABC}\hat{L}^{C}\,,&
    \left[S_A,\hat{P}_B\right]&=\epsilon_{ABC}\hat{L}^{C}\,, \notag \\
    \left[\hat{L}_{A},\hat{P}_{B}\right]&=-\Lambda\epsilon_{ABC}S^{C}\,, \label{eMB}
\end{align}
with $\Lambda$ being defined as in \eqref{Lambda}. Let us note that when $p=q=0$, the enhanced MB algebra \eqref{EnhMB} reduces to the coadjoint Poincaré algebra \cite{Barducci:2019jhj,Barducci:2020blv,Bergshoeff:2020fiz}. Furthermore, when $q=0$ and $p=1/\ell^{2}$, it reproduces the coadjoint AdS algebra. On the other hand, when $p=0$ and $q=-2/\ell$ the enhanced teleparallel algebra introduced in \cite{Concha:2022you} is recovered. Analogously to the cases studied in \cite{Bergshoeff:2020fiz,Concha:2022you}, to obtain a non-degenerate invariant tensor in the non-relativistic limit, it is also necessary to include $\mathfrak{u}(1)$ generators, $Y_{1}$ and $Y_{2}$, as it was considered in the previous section. At the relativistic level, the non-vanishing components of the invariant bilinear form for the enlarged enhanced MB algebra are given by \eqref{invtensor}, \eqref{intenu} , and those components involving the additional generators $( S_A, L_A )$:
\begin{equation}\label{invtensor3}
    \langle J_{A} S_{B} \rangle = \beta_{2} \,\eta_{A B}\,,\qquad \langle J_{A} L_{B} \rangle = \beta_{1} \,\eta_{A B}\,,\qquad\langle P_{A} S_{B} \rangle = \beta_{1} \,\eta_{A B}\,,\qquad \langle P_{A} L_{B} \rangle = \left(p \beta_{2}+q\beta_{1}\right)\eta_{A B}\,,
\end{equation}
where $\beta_1$ and $\beta_2$ are arbitrary constants. Then, the enhancement of the MB algebra does not modify the original MB Lagrangian \eqref{MB} but add new contributions along $\beta_1$ and $\beta_2$ which will be crucial to elucidate a Newtonian version of the MB gravity theory.


\subsection{Newtonian MB algebra}

A non-relativistic version of an enlargement and enhancement of the MB algebra can be obtained by applying an Inönü-Wigner contraction to \eqref{algebra01} and \eqref{EnhMB}. To apply the contraction we have to first express the relativistic generators as a linear combination of the non-relativistic ones through a dimensionless parameter $\xi$ as follows:
\begin{align}
J_{0} &=\frac{J}{2}  -\xi ^{4}Z\,,& J_{a}&=\frac{
\xi }{2}G_{a}-\frac{\xi ^{3}}{2}B_{a}\,,  \notag \\
P_{0} &=\frac{H}{2}  -\xi ^{4}Y\,,& P_{a}&=\frac{
\xi }{2}P_{a}-\frac{\xi ^{3}}{2}L_{a}\,,  \notag \\
S_{0} &=-\xi ^{2}S-\xi^{4}Z\,,& S_{a}&=-\xi G_{a}-\xi ^{3}B_{a}\,,  \notag \\
L_{0} &=-\xi ^{2}M-\xi^{4}Y\,,& L_{a}&=-\xi T_{a}-\xi ^{3}L_{a}\,,  \notag \\
Y_{1}&= \frac{J}{2}  +\xi ^{4}Z\,, & Y_{2}&=\frac{H}{2}  +\xi ^{4}Y\,. \label{redefMB}
\end{align}

Then, in the limit $\xi\rightarrow\infty$, the non-relativistic generators $\left \lbrace J, H, G_{a}, P_{a}, S, M, B_{a}, L_{a}, Y, Z\right\rbrace$, satisfy the commutation relations of the extended Newtonian algebra:
\begin{align}
\left[ J,G_{a}\right] &=\epsilon _{ab}G_{b}\,, &
\left[ G_{a},G_{b}\right] &=-\epsilon _{ab}S\,, &
\left[ H,G_{a}\right] &=\epsilon _{ab}P_{b}\,,  \notag
\\
\left[ J,P_{a}\right] &=\epsilon _{ab}P_{b}\,,&
\left[ G_{a},P_{b}\right] &=-\epsilon _{ab}M\,,&
\left[ H,B_{a}\right] &=\epsilon _{ab}L_{b}\,,  \notag
\\
\left[ J,B_{a}\right] &=\epsilon _{ab}B_{b}\,,&
\left[ G_{a},B_{b}\right] &=-\epsilon _{ab}Z\,,&
\left[ S,G_{a}\right] &=\epsilon _{ab}B_{b}\,,
\notag \\
\left[ J,L_{a}\right] &=\epsilon _{ab}L_{b}\,,&
\left[ G_{a},L_{b}\right] &=-\epsilon _{ab}Y\,,&
\left[ S,P_{a}\right] &=\epsilon _{ab}L_{b}\,,
\notag \\
\left[ M,G_{a}\right] &=\epsilon _{ab}L_{b}\,,&
\left[ P_{a},B_{b}\right] &=-\epsilon _{ab}Y\,, \label{EN}
\end{align}
along with 
\begin{align}
\left[ H,P_{a}\right] &=\epsilon _{ab}\left(p G_{b}+q P_{b}\right)\,, &
\left[ P_{a},P_{b}\right] &=-\epsilon _{ab}(p S+ q M)\,, &
\left[ H,L_{a}\right] &=\epsilon _{ab}(p B_{b}+q L_{b}) \,,  \notag
\\
\left[ M,P_{a}\right] &=\epsilon _{ab}(p B_{b}+q L_{b})\,,&
\left[ P_{a},L_{b}\right] &=-\epsilon _{ab}(p Z+ q Y)\,. \label{MBEN}
\end{align}
This new non-relativistic symmetry is denoted as the Newtonian MB (NMB) algebra. Such symmetry, unlike the Newtonian one introduced in \cite{Hansen:2019pkl}, is characterized by the presence of two central charges $Z$ and $Y$ which, as we shall see, are required for having non-degenerate invariant bilinear trace.
As in the previous section, different Newtonian type algebras can be derived from the NMB algebra. Indeed, the Extended Newtonian algebra \cite{Ozdemir:2019orp}, the Newton-Hooke version of the Newtonian algebra \cite{Concha:2019dqs,Gomis:2019nih,Bergshoeff:2020fiz} and the torsional extended Newtonian (TEN) algebra \cite{Concha:2022you} are obtained once we fix the $(p,q)$ parameters as shown in Table \ref{Table2}.
\begin{table}[!h]
\centering
    \begin{tabular}{|c|c|c|}
\hline
Newtonian type algebra  & $p $ & $q$ \\ \hline\hline
TEN algebra& 
   $0$ & $-2/\ell$ \\
   Newton-Hooke-Newtonian algebra &
   $1/\ell^{2}$ & $0$ \\ 
Extended Newtonian algebra & $0$ & $0$\\  \hline
\end{tabular}\notag
\caption{Newtonian symmetries for different values of $p$ and $q$ in the NMB algebra.}
\label{Table2}
\end{table}

 \subsection{Newtonian MB Chern-Simons gravity action}
For the construction of the CS Lagrangian it is required the invariant tensor for the NMB algebra. It is possible to show that the non-vanishing components of the invariant tensor are: 
\begin{eqnarray}
\langle S S \rangle&=&\langle J Z \rangle=-\tilde{\beta}_2\,, \notag \\
\langle G_a B_b \rangle&=&\tilde{\beta}_2\delta_{ab}\,, \notag \\
\langle M S \rangle&=&\langle H Z\rangle=\langle J Y \rangle=-\tilde{\beta}_1\,, \notag \\
\langle P_a B_b \rangle&=&\langle G_a L_b \rangle=\tilde{\beta}_1 \delta_{ab}\,, \notag \\
\langle H Y\rangle&=&\langle M M \rangle=-\left(p\tilde{\beta}_2+ q\tilde{\beta}_1\right)\,, \notag \\
\langle P_{a} L_{b}\rangle &=&
\left(p\tilde{\beta}_2+q\tilde{\beta}_1\right)\delta_{ab}\,. \label{IT4}
\end{eqnarray}
which can be obtained from the relativistic components \eqref{invtensor}, \eqref{intenu} and \eqref{invtensor3} by applying the limit $\xi\rightarrow\infty$ after considering the contraction of the generators \eqref{redefMB} and the rescaling of the relativistic parameters as
\begin{eqnarray}
\sigma_{2}=\beta_{2}=-\tilde{\beta}_2\xi^{4}\,, \qquad \qquad \sigma_{1}=\beta_{1}=-\tilde{\beta}_1\xi^{4}\,. \label{REDEF3}
\end{eqnarray}

Furthermore, the NMB algebra also admits the bilinear invariant trace for the non-relativistic MB algebra defined in the previous section, namely
\begin{align}
\langle J S\rangle &=-\tilde{\sigma}_{2}\,,\notag\\
\langle G_{a} G_{b} \rangle &=\tilde{\sigma}_{2}\delta_{ab}\,, \notag\\
\langle G_{a} P_{b} \rangle &=\tilde{\sigma}_{1}\delta_{ab}\,, \notag\\
\langle H S \rangle &=\langle M J \rangle=-\tilde{\sigma}_{1}\,, \notag\\
\langle P_{a} P_{b} \rangle &=\left(p\tilde{\sigma}_{2}+q\tilde{\sigma}_{1}\right)\delta_{ab}\,, \notag\\
\langle H M \rangle &=-\left(p\tilde{\sigma}_{2}+q\tilde{\sigma}_{1}\right)\,,\label{ITNRMB2}
\end{align}
where the relativistic parameters obey $\sigma_{2}=\sigma_{1}=0$ along the following rescaling:
\begin{equation}
\beta_{2}=-\tilde{\sigma}_{2}\xi^{2}\,, \qquad \qquad \beta_{1}=-\tilde{\sigma}_{1} \xi^{2} \,.  \label{alphas}
\end{equation}

Let us note that the components of the invariant tensor proportional to $\tilde{\sigma}$'s are degenerate for the whole NMB algebra although they define a non-degenerate invariant trace for the  NR MB algebra obtained in the previous section. The non-degeneracy requires to consider the invariant tensor given by \eqref{IT4} or to consider both contributions proportional to $\tilde{\sigma}$'s and $\tilde{\beta}$'s. For completeness, we shall consider the complete set of non-vanishing components of the invariant tensor keeping in mind that two inequivalent limits at the level of the relativistic CS constants are considered for obtaining the NR invariant tensor \eqref{IT4} and \eqref{ITNRMB2}.

The one-form gauge connection for the NMB algebra reads
\begin{equation}
    A=\tau H+e^{a}P_{a}+\omega J+\omega^{a}G_{a}+mM+sS+l^{a}L_{a}+b^{a}B_{a}+yY+zZ\,. \label{oneformNMB}
\end{equation}
The curvature two-form $F=dA+\frac{1}{2}\left[ A,A\right] $ is given by 
\begin{eqnarray}
F&=&R\left(\tau\right) H+R^{a}\left(e^{b}\right)P_{a}+R\left(\omega\right) J+R^{a}\left(\omega^{b}\right)G_{a}+R\left(m\right)M+R\left(s\right)S+R^{a}\left(l^{b}\right)L_{a} \notag\\
&&+R^{a}\left(b^{b}\right)B_{a}+R(y)Y+R(z)Z\,, \label{curvNMB}
\end{eqnarray}
where the components are explicitly given by:
\begin{eqnarray}
R(\omega ) &=&d\omega\,,  \notag\\
R(s) &=&ds+\frac{1}{2}\epsilon ^{ac}\omega _{a}\omega _{c}+\frac{p}{2}\epsilon ^{ac}e_{a}e_{c}\,, \notag\\
R(z) &=&dz+\epsilon ^{ac}\omega _{a}b_{c}+p\epsilon ^{ac}e_{a}l_{c}\,, \notag \\
R(\tau ) &=&d\tau \,, \notag \\
R(m) &=&dm+\epsilon ^{ac}\omega _{a}e_{c}+\frac{q}{2}\epsilon ^{ac}e_{a}e_{c}\,, \notag
\\
R(y) &=&dy+\epsilon ^{ac}\omega _{a}l_{c}+\epsilon ^{ac}b_{a}e_{c}+q\epsilon ^{ac}e_{a}l_{c}\,, \notag \\
R^{a}(\omega ^{b}) &=&d\omega ^{a}+\epsilon ^{ac}\omega \omega _{c}+p\epsilon ^{ac}\tau e_{c}\,, \notag \\
R^{a}(b^{b}) &=&db^{a}+\epsilon ^{ac}\omega b_{c}+\epsilon ^{ac}s\omega _{c}+p
\epsilon ^{ac}\tau l_{c}+p\epsilon ^{ac}me_{c}\,, \notag
\\
R^{a}(e^{b}) &=&de^{a}+\epsilon ^{ac}\omega e_{c}+\epsilon ^{ac}\tau \omega
_{c}+q\epsilon ^{ac}\tau e_{c}\,, \notag \\
R^{a}(l^{b}) &=&dl^{a}+\epsilon ^{ac}\omega l_{c}+\epsilon
^{ac}se_{c}+\epsilon ^{ac}\tau b_{c}+\epsilon ^{ac}m\omega _{c}+q
\epsilon ^{ac}\tau l_{c}+q\epsilon ^{ac}me_{c}\,. \label{ccurv}
\end{eqnarray}
From the previous expressions, we can see that when $p=q=0$, the curvature two-forms of the extended Newtonian are recovered. On the other hand, when we fix $q=0$ and $p=1/\ell^{2}$, the curvatures reduce to those of the Newton Hooke version of the Newtonian algebra. Similarly, when $p=0$ and $q=-2/\ell$, the TEN curvatures are obtained. The same analysis is valid at the level of the invariant tensors. Let us note that if we consider the contraction at the level of the CS Lagrangian based on the enhancement and enlargement of the MB algebra, the resulting non-relativistic Lagrangian will depend on the choice of the rescaling of the arbitrary constants, namely, \eqref{REDEF3} or \eqref{alphas}. To construct the most general and non-degenerate Lagrangian for the NMB algebra, we will consider both families of invariant tensors.

A CS Lagrangian based on the NMB algebra \eqref{EN} and \eqref{MBEN} is constructed considering the gauge connection one-form \eqref{oneformNMB} and the non-vanishing components of the invariant tensor \eqref{IT4} and \eqref{ITNRMB2} in the CS Lagrangian \eqref{CS},
\begin{eqnarray}
    L &=&L_{\text{NRMB}} +L_{\text{NMB}}\,, 
\end{eqnarray}
where $ L_{\text{NMB}}$ is the non-relativistic Lagrangian \eqref{NRCS} obtain previously and $L_{\text{NMB}}$ is given by
\begin{eqnarray}
L_{\text{NMB}} &=& -\tilde{\beta}_{0}\left(\epsilon^{ac}\tau e_{a}l_{c}+\epsilon^{ac}m e_{a}e_{c}\right)+\tilde{\beta}_{1}\left[
2 e_{a}\hat{R}^{a}\left(b ^{b}\right)+2l_{a}\hat{R}^{a}\left(\omega^{b}\right)-2\tau \hat{R}(z)-2 m \hat{R}(s)\right.\notag \\ 
&& \left.-2 yR(\omega)\right] +\tilde{\beta}_{2}\left[ \omega _{a}\hat{R}^{a}\left(b
^{b}\right)+b_{a}\hat{R}^{a}\left(\omega
^{b}\right)-2zR\left( \omega \right)-sds \right]+\tilde{\beta}_{3}\left[
e_{a}\hat{R}^{a}\left(l ^{b}\right)+l_{a}\hat{R}^{a}\left(e ^{b}\right)\right. \notag\\
&& \left. -m \hat{R}(m)-\tau R(y)-y R(\tau)\right] \,.  \label{NMBCS}
\end{eqnarray}
where we have defined 
\begin{align}
    \hat{R}^{a}\left( b ^{b}\right)  &= db ^{a}+\epsilon^{ac}\omega b_{c}+\epsilon^{ac}s \omega_{c}\,, \notag \\
     \hat{R}^{a}\left( l^{b}\right) &= d l^a + \epsilon ^{ac}\omega l_{c}+\epsilon ^{ac}s e_{c}+\epsilon ^{ac}\tau b_{c}\,, \notag\\
        \hat{R}\left( y\right) &= dy+ \epsilon ^{ac} \omega_{a}l_{c}+ \epsilon ^{ac} b_{a}e_{c}\,, \notag\\
    \hat{R}\left( z\right) &= dz+ \epsilon ^{ac} \omega_{a}b_{c}\,, 
\end{align}
and
\begin{align}
\tilde{\beta}_{3}&=p \tilde{\beta}_{2}+q \tilde{\beta}_{1}\,, \notag \\
    \tilde{\beta}_{0}&=p\tilde{\beta}_{1}+q \tilde{\beta}_{3}\,. \label{redef2}
\end{align}
Let us note that the term proportional to $\tilde{\beta_1}$ is the extended Newtonian Lagrangian introduced in \cite{Ozdemir:2019orp}. The cosmological term appears along $\tilde{\beta_0}$. On the other hand, the term along $\tilde{\beta_2}$ can be seen as the Newtonian version of the CS gravitational term. The torsional term of the Newtonian MB gravity theory appears along $\tilde{\beta_3}$. It also turns out appealing that
the diverse Newtonian gravity Lagrangians known in the literature \cite{Ozdemir:2019orp,Concha:2019dqs,Gomis:2019nih,Bergshoeff:2020fiz,Concha:2022you}, appear by fixing the $\left(p,q\right)$ parameters as in Table \ref{Table2}.

In presence of non-degenerate invariant bilinear form the field equations reduce to the vanishing of the curvature two-forms. Here, the equations of motion of the NMB gravity theory derived from \eqref{NMBCS} read
\begin{eqnarray}
    \delta\omega^{a}&:& \qquad \tilde{\beta}_{1}R^{a}\left(s^{b}\right)+\tilde{\beta}_{2}R^{a}\left(b^{b}\right)=0\,, \notag \\
    \delta\omega&:& \qquad \tilde{\beta}_{1}R\left(y\right)+\tilde{\beta}_{2}R\left(z\right)=0\,, \notag \\
    \delta e^{a}&:& \qquad \tilde{\beta}_{1}\left[R^{a}\left(b^{b}\right)+qR^{a}\left(l^{b}\right)\right]+\tilde{\beta}_{2}\,pR^{a}\left(l^{b}\right)=0\,, \notag \\
    \delta\tau&:& \qquad \tilde{\beta}_{1}\left[R\left(z\right)+qR\left(y\right)\right]+\tilde{\beta}_{2} \,pR\left(y\right)=0\,, \notag \\
    \delta s&:& \qquad \tilde{\beta}_{1}R\left(m\right)+\tilde{\beta}_{2}R\left(s\right)=0\,, \notag \\
    \delta m&:& \qquad \tilde{\beta}_{1}\left[R\left(s\right)+qR\left(m\right)\right]+\tilde{\beta}_{2}\,pR\left(m\right)=0\,, \notag \\
    \delta z&:& \qquad \tilde{\beta}_{1} R\left(\tau\right) + \tilde{\beta}_{2} R\left(\omega\right)=0\,,\notag \\
    \delta b^{a}&:& \qquad \tilde{\beta}_{1}R^{a}\left(e^{b}\right)+\tilde{\beta}_{2}\,R^{a}\left(\omega^{b}\right)=0\,, \notag \\
    \delta y&:& \qquad \tilde{\beta}_{1}\left[ R\left(\omega\right)+qR\left(\tau\right)\right] + \tilde{\beta}_{2}\,p R\left(\tau\right)=0\,, \notag \\
    \delta l^{a}&:& \qquad \tilde{\beta}_{1}\left[R^{a}\left(\omega^{b}\right)+qR^{a}\left(e^{b}\right)\right]+\tilde{\beta}_{2}\,pR^{a}\left(e^{b}\right)=0\,,
    \label{eom2}
\end{eqnarray}
where we have used \eqref{redef2}. The non-degeneracy of the invariant tensor implies that $\tilde{\beta}_{1}^{2}-\tilde{\beta}_2\tilde{\beta}_{3}\neq0$ and $\tilde{\beta_1}\neq0$ which ensures the vanishing of the NMB curvatures \eqref{ccurv}. One can note that, as in the NR MB case studied previously, only the spatial component of the torsion $\hat{R}^{a}\left(e^{b}\right)=-q\epsilon^{ac}\tau e_c$ remains turned on.

\section{Conclusions}\label{sec5}

In this work, we presented the NR regime of the MB gravity model. To this end, we have applied an NR limit to the so-called MB algebra enlarged with two $\mathfrak{u}\left(1\right)$ generators. Such enlargement ensures the presence of central charges allowing to define a non-degenerate invariant bilinear trace. As in the relativistic MB gravity, the NR counterpart contains a source for both, constant torsion and constant curvature measured by the parameters $q$ and $p$, respectively. We were able to make contact with different NR gravity models defined in three spacetime dimensions by considering specific values of the $\left(p,q\right)$ parameters. Subsequently, we extended our results to the Newtonian realm by considering the NR limit to an enhanced MB algebra enlarged with two $\mathfrak{u}\left(1\right)$ generators.

The results obtained here could bring valuable information about the role of torsion in the NR regime from the MB formalism. In particular, both NR and Newtonian versions of the MB gravity are characterized by containing sources for the diverse components of the curvature two-form $F$ in terms of the $\left(p,q\right)$ parameters. However, the time-component of the torsion remains equal to zero in the NR and Newtonian limit for any value of the $\left(p,q\right)$ parameters. It would be then interesting to study if our model can be related to the torsional Newton-Cartan gravity theory \cite{Bergshoeff:2015ija,Bergshoeff:2017dqq,VandenBleeken:2017rij} in which the non-zero torsion condition implies the presence of a non-vanishing time-like torsion. As it was noticed in \cite{Concha:2021llq}, having a spatial-component of the torsion in the NR teleparallel gravity would imply, at the level of the boost behavior, introducing a non-zero time-like torsion as well. Let us note that a non-vanishing time-like torsion condition in an NR environment has first been encountered in Lifshitz holography context \cite{Christensen:2013lma} and Quantum Hall effect \cite{Geracie:2015dea}.

One could go further in the study of the NR version of an MB gravity model. It would be interesting to include supersymmetry and higher-spin gauge fields in our NR model. Despite the numerous applications of both supergravity and higher-spin gravity in the relativistic context, NR supergravity and NR gravity coupled to higher spin has just been recently approached in \cite{Andringa:2013mma,Bergshoeff:2015ija,Bergshoeff:2016lwr,Ozdemir:2019orp,Concha:2020eam} and \cite{Bergshoeff:2016soe,Concha:2022muu,Caroca:2022byi}, respectively. On the other hand, the supersymmetric extension of the MB gravity has been explored in \cite{Giacomini:2006dr,Cvetkovic:2007sr,Caroca:2021njq}. Nevertheless, the NR limit of MB supergravity cannot naively be applied due to the appearance of degeneracy. One way to circumvent the difficulty encountered in the NR contraction process is to consider the Lie algebra expansion method \cite{Hatsuda:2001pp,deAzcarraga:2002xi,Izaurieta:2006zz,deAzcarraga:2007et}. As it was noticed in \cite{Gomis:2019nih}, the semigroup expansion method \cite{Izaurieta:2006zz} offers us a straightforward mechanism to derive a non-degenerate NR algebra from a relativistic one for a particular semigroup $S$. Then, following the procedure employed in the presence of supersymmetry \cite{deAzcarraga:2019mdn,Ozdemir:2019tby,Concha:2019mxx,Concha:2020tqx,Concha:2021jos}, one could elucidate the corresponding supersymmetric extension of our results. One could expect that the extended Bargmann supergravity \cite{Bergshoeff:2016lwr}, the extended Newton-Hooke \cite{Ozdemir:2019tby} and the recent NR teleparallel supergravity \cite{Concha:2021llq} appear for particular values of the $\left(p,q\right)$ parameters. In the higher-spin case, one could start by exploring the NR limit of the spin-3 MB gravity theory studied in \cite{Peleteiro:2020ubv} and check if the spin-3 extended Bargmann gravity introduced in \cite{Bergshoeff:2016soe,Concha:2022muu} appears as a particular subcase.

\section*{Acknowledgments}

This work was funded by the National Agency for Research and Development ANID - SIA grant No. SA77210097 and FONDECYT grants No. 1211077, 1231133, 11220328 and 11220486.  P.C. and E.R. would like to thank to the Dirección de Investigación and Vice-rectoría de Investigación of the Universidad Católica de la Santísima Concepción, Chile, for their constant support.


\bibliographystyle{fullsort.bst}
 
\bibliography{Non_relativistic_MB}

\end{document}